\documentclass{mn2e}
\usepackage{epsfig}
\usepackage{times}
\usepackage{hyperref}
\begin{document}

\title[DG CVn] 
{A prompt radio transient associated with a gamma-ray superflare from the young M dwarf binary DG CVn}
%{Prompt, bright radio flaring associated with a superoutburst from the young M dwarf binary DG CVn}
%{A bright radio flare associated with the gamma ray superoutburst of the nearby binary DG CVn}

\author[Fender]
       {R.P. Fender$^1$\thanks{email: rob.fender@astro.ox.ac.uk}, G.E. Anderson$^1$, R. Osten$^2$, T. Staley$^{1,3}$, 
C. Rumsey$^4$, K. Grainge$^5$, \newauthor R.D.E. Saunders$^4$
\\
$^1$Astrophysics, Department of Physics, University of Oxford, Keble Road, Oxford OX1 3RH\\
$^2$Space Telescope Science Institute, 3700 San Martin Drive, Baltimore, MD 21218, USA\\
$^3$Physics \& Astronomy, University of Southampton, Southampton SO17 1BJ\\ 
$^4$Astrophysics Group, Cavendish Laboratory, JJ Thomson Avenue, Cambridge CB3 0HE, UK\\
$^5$Jodrell Bank Centre for Astrophysics, School of Physics and Astronomy, The University of Manchester, Manchester M13 9PL
\\}
\maketitle
\begin{abstract}
On 2014 April 23, the {\em Swift} satellite detected a gamma-ray superflare from the nearby star system DG CVn. This system comprises a M-dwarf 
binary with extreme properties: it is very young and at least one of the components is a very rapid rotator. The gamma-ray superflare is one of only a handful detected by {\em Swift} in a decade. As part of our AMI-LA Rapid Response Mode, ALARRM, we automatically slewed to this target, were taking data at 15 GHz within six minutes of the burst, and detected a bright ($\sim 100$ mJy) radio flare. This is the earliest detection of bright, prompt, radio emission from a high energy transient ever made with a radio telescope, and is possibly the most luminous incoherent radio flare ever observed from a red dwarf star. An additional bright radio flare, peaking at around 90 mJy, occurred around one day later, and there may have been further events between 0.1--1 days when we had no radio coverage. The source subsequently returned to a quiescent level of 2--3 mJy on a timescale of about 4 days. Although radio emission is known to be associated with active stars, this is the first detection of large radio flares associated with a gamma ray superflare, and demonstrates both feasibility and scientific importance of rapid response modes on radio telescopes.
\end{abstract}

\section{Introduction}

Variable and transient radio emission is a strong indicator of unusual and/or energetic activity in an astrophysical object. At its 
most extreme it can be a marker for the feedback of vast amounts of kinetic energy into the local medium (in, for example, gamma-ray 
bursts, supernovae, active galactic nuclei, black holes and neutron stars), or of coherent processes representing some of the highest 
energy densities in the universe (pulsars, fast radio bursts). For this reason, searching for and finding radio variables has become a 
key goal for the current and next generation of radio telescopes (e.g. LOFAR, MWA, MeerKAT, ASKAP, LWA) {\em en route} to the 
Square Kilometre Array (see e.g. Fender \& Bell 2011; Macquart 2014 and references therein).

Magnetic reconnetion-induced flaring variability from stars in the cool half of the HR diagram is expected due to the presence and action of magnetic fields in the stars' outer atmospheres; while the Sun is the best studied flaring star it is the nearly- to fully-convective M dwarfs which have displayed some of the most extreme outbursts. A flare event encompasses a host of physical processes, including particle acceleration and plasma heating, and involves all layers of the stellar atmosphere. Microwave stellar radio emission associated with flares is typically produced as the result of particles accelerated during the flare process, and emitting gyrosynchrotron radiation while they are trapped in the magnetic flaring coronal loop.  The timescale of the event relates to the timescales for the acceleration of electrons, their injection into the coronal loop,
and the trapping time.  Because of the high space density of M dwarfs they are expected to be detected as transients in many regions of the electromagnetic spectrum, from
radio to optical to X-ray wavelengths.

DG CVn (GJ 3789) is a nearby, but rather extreme, stellar system. 
It is listed in the Washington Double Star Catalog as a binary of M dwarfs with separation $\sim$0.2"; recent trigonometric parallax measurements of Riedel et al. (2014) place the system at a distance of 18 pc, with an estimated age of only 30 MY. The system has fast rotation -- Mohanty \& Basri (2003) list a $v\sin i$ of $\approx$ 50 km s$^{-1}$, has been detected in previous radio surveys at a level of a few mJy (Bower et al. 2009), and is detected in the ROSAT All-Sky Survey Bright Source Catalog. Despite these fascinating properties, the system has to date eluded detailed study of its flaring variability.

\begin{figure*}
\centerline{\epsfig{file=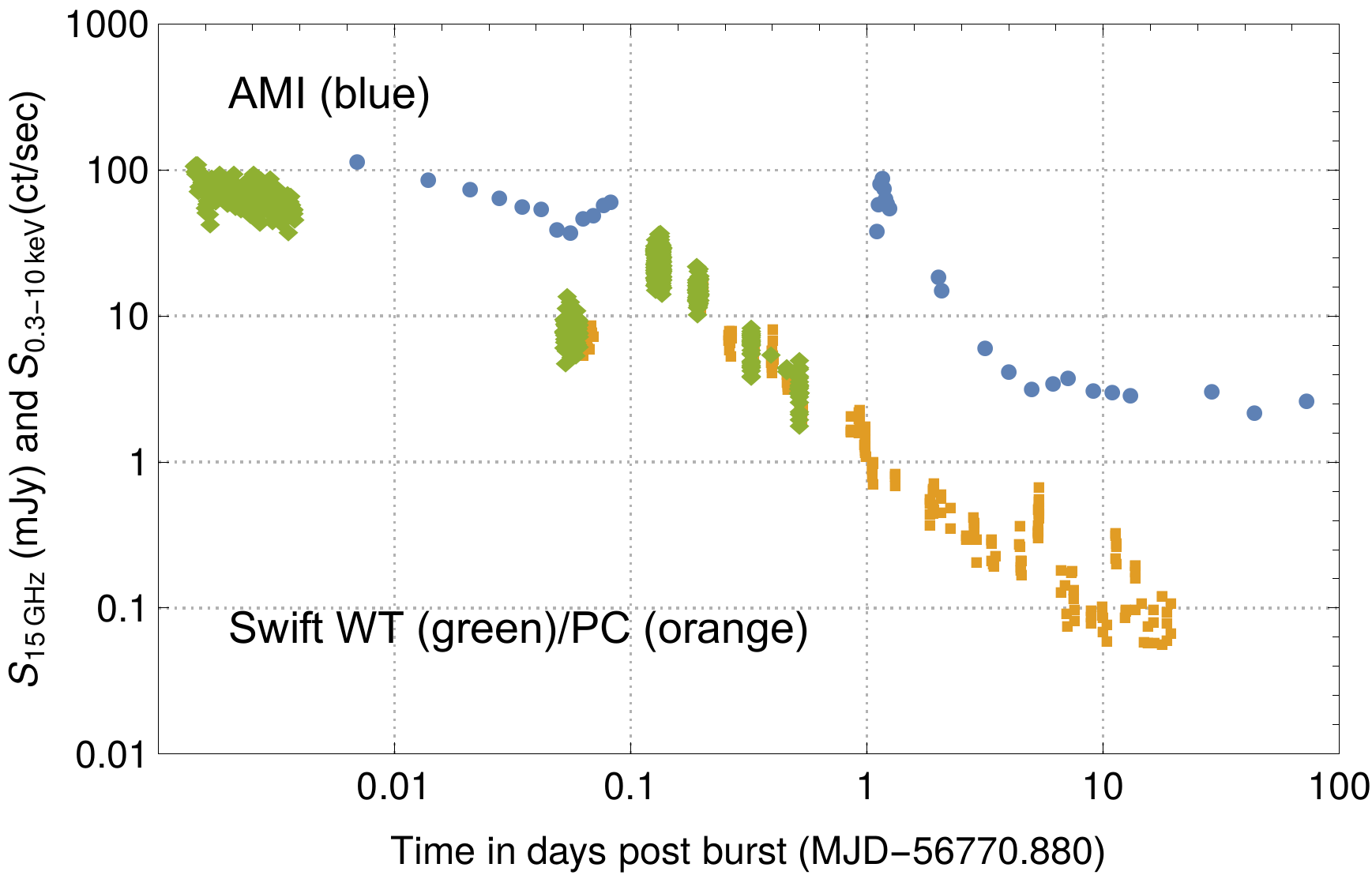, angle=0, width=14.5cm}}
\caption{AMI-LA 13--18 GHz (blue circles) and Swift WT/PC (green diamonds/orange squares) 0.3--10 keV light curves on a logarithmic scale. 
For clarity, error bars are not plotted but for both radio and X-ray data are typically $\leq 15$\%, and are plotted in Fig 2, which presents the flare on a linear scale.
In X-rays the source was brightest at the first measurement, two minutes after the initial trigger, and then declined for around the first hour, rebrightening somewhere between 0.075 and 0.125 days. The radio flux behaved similarly, with a bright, strong detection in the first measurement at six minutes, followed by a decline and subsequent rebrightening. A second, clearly resolved, radio flare (revealed more clearly in Fig 2) occurred at around 1.1 days. By about four days the radio flux had settled down to a `quiescent' level of a few mJy. 
}
\end{figure*}

On 2014 April 23, Drake et al. (2014) reported a hard X-ray superflare from DG CVn, detected with the Burst Alert Telescope (BAT) onboard the {\em Swift} satellite, with a peak 15--50  keV flux of $\sim$ 300 mCrab. This was bright enough to trigger {\em Swift} and cause it to slew automatically to the source  (as if it were a GRB). This stellar superflare is one of only a handful ($\leq 6$) which have been bright enough to trigger {\em Swift}
in such a fashion, and the previous ones (Osten et al. 2007, Osten et al. 2010) have revealed the extreme end of magnetic reconnection in normal nondegenerate stars. 

As part of our robotic response to {\em Swift} triggers with the AMI-LA radio telescope, a radio observation of this event was scheduled within a minute of receipt of the VOEvent from GSFC, and data were first taken 5 minutes and 18 seconds after the {\em Swift} trigger. Radio monitoring proceeded for $\approx$ 73 days post-trigger, enabling a long-term perspective on the radio variability history of the system. We have detected bright peaks (flux density at 15 GHz $\geq 100$ mJy) and rapid evolution in these AMI observations. This is the first time strong radio transients have been associated with such an X-ray superflare, and in this paper we report the details of this event. 

\section{Observations}

\subsection{AMI-LA / ALARRM}

The Arcminute Microkelvin Imager Large Array (AMI-LA; Zwart et al. 2008) is a compact array of eight dishes operating in the 14--18 GHz frequency range. Since March 2012 it has been operating in an automatic response mode to well-localised {\em Swift} GRBs (Staley et al. 2013; Anderson et al. 2014), routinely observing events within minutes, delivering the fastest regular radio response to high energy events (but see also Palaniswamy et al. 2014 for similar efforts with a single dish).

In 2014, we have adjusted this programme to respond to all {\em Swift} triggers, and not to cancel (sometimes before any data are taken) observations of previously-known objects which trigger {\em Swift} from time to time, such as flare stars or X-ray binaries. This approach widens the scientific scope of the programme, and allows us to generate an unbiased sample of prompt radio observations of flaring gamma ray sources, whatever their physical origin. As part of this expansion of the programme, we rename the project as the {\em AMI-LA Rapid Response Mode}, ALARRM. This is at present the only continuously operating radio facility with a robotic transient response mode.

The data presented in this paper were initially obtained as part of the ALARRM programme, in response to the {\em Swift} trigger on DG DVn and further observations were scheduled manually when it became apparent that there was a bright radio event occuring. 
Table 1 presents the timeline of events from trigger to first radio data, involving the ingestion of the NASA VOEvent by the 4 PI SKY node and subsequent robotic trigger request to AMI. We are currently working on further adjustments to the ALARRM system which should reduce the time delay before the first AMI data are taken.
The radio data are presented in Table 2 and Figs 1 and 2, where flare time is taken as MJD 56770.880.

The first, robotically triggered, observation was severely affected by heavy rain so standard AMI reduction techniques resulted in a large amount of the data being flagged. 
We therefore applied a
less stringent flagging regime and calibrated the flux using an AMI
observation of 3C286 conducted on 2014 April 24 under similar
observing conditions. Including data from channels 3-8, the resulting
AMI dataset was $<$ 30\% flagged allowing us to divide the 2
hour observation into twelve 10 minute images where DG CVn is
clearly seen to be recovering from a very early time radio flare. The
mean flux level of the phase calibrator J1332+2949 during the 2014 April 23 AMI observation was 10\% brighter than its mean flux level during better weather conditions.
We therefore assumed a 10\% flux calibration error for all 12 measurements taken from this early time AMI observation of DG CVn, which are listed in Table 2.

One day later we observed for four hours, and split this observation into eight separate 30 minute imges. We clearly see a rise and fall in the 15 GHz emission, peaking at around 1.1 days post-burst. Subsequent observations were made at approximately daily cadence with 1--3 hr integration times. After two weeks we made a further three observations at weekly--monthly intervals in order to track the longer-term behaviour.

Aside from the first observation, data reduction of the DG CVn AMI observations was conducted with the AMI \texttt{REDUCE} software using the source J1332+2949 for phase calibration (Perrott et al. 2013). No further flagging was performed after this automatic step. Imaging was conducted using the Common Astronomy Software Application package (\texttt{CASA;} Jaeger 2008). The integrated fluxes quoted in Table 2 were measured using the \texttt{Miriad} (Sault, Teuben \& Wright 1995) task \texttt{imfit} assuming a Gaussian model. All flux errors correspond to the quadratic sum of the image rms and the 5 percent flux calibration error of AMI (Perrott et al. 2013). The fluxes were then primary beam corrected as each observation’s pointing centre was 1.4 arcmin offset from the position of DG CVn. We note that we were not confident in our ability to measure the in-band radio spectral index from these observations, in part because the source was significantly off-axis, but that the situation should improve with the implementation of a new digital correlator within the next 12 months.
For further details on the reduction and analysis we performed on the AMI observations please see Anderson et al. (2014). 

\begin{table}
\caption{Timeline of the triggering of the ALARRM system by the {\em Swift} trigger on DG CVn}
\begin{tabular}{rl}
Time & Event\\
\hline
UT 23 Apr 21:08:25 & Initial Swift BAT timestamp \\
& (VOEvent, marked as GRB) \\
            21:08:29 & 4 Pi Sky Bot activated \\ 
            21:08:30 & AMI receives request email, ALARRM on\\
            21:08:35 & AMI starts slew to phase cal.\\ 
            21:10:33 & Swift DG CVn observation starts \\  
	    21:10:34 & AMI on phase cal., waits \\
            21:12.26 & AMI phase cal. observation starts \\
            21:14:06 & phase cal. obs complete \\
& starts slew to DG CVn		\\
            21:14:21 & AMI DG CVn observation starts\\
\hline
\end{tabular}
\end{table}

We do not attempt in this paper to perform anywhere near a full modelling analysis of the radio data in the context of models for flare star outbursts, nor of its connection to data obtained at other wavelengths (apart from {\em Swift}). However, we do encourage the community to do so, by comparison with their own data and models, and so we present all of our measurements in Table 2. Further details, if required, may be obtained by contacting the authors of the paper. 

\begin{table}
\caption{AMI-LA observations of DG CVn at a central frequency of 15.7 GHz. Observation times correspond to the midpoint of the observation.}
\begin{tabular}{rcc}
MJD-56770.88 & duration (hr) & $F_{\rm 15 GHz}$ \\ 
\hline
0.007 & 0.17 & $116.58 \pm 12.85$ \\
0.014 & 0.17 & $87.6 \pm 9.92$ \\
0.021 & 0.17 & $75.75 \pm 8.54$ \\
0.028 & 0.17 & $66.45 \pm 7.61$ \\
0.035 & 0.17 & $57.28 \pm 6.46$ \\
0.042 & 0.17 & $55.61 \pm 6.09$ \\
0.049 & 0.17 & $40.24 \pm 4.79$ \\
0.056 & 0.17 & $38.3 \pm 4.41$ \\
0.063 & 0.17 & $47.51 \pm 5.3$ \\
0.070 & 0.17 & $50.19 \pm 5.48$ \\
0.077 & 0.17 & $59.0 \pm 6.35$ \\
0.083 & 0.17 & $62.16 \pm 6.57$ \\
1.104 & 0.5 & $38.94 \pm 2.12$ \\
1.125 & 0.5 & $59.81 \pm 3.19$ \\
1.146 & 0.5 & $82.51 \pm 4.39$ \\
1.166 & 0.5 & $90.66 \pm 4.82$ \\
1.187 & 0.5 & $76.29 \pm 4.0$ \\
1.208 & 0.5 & $65.52 \pm 3.57$ \\
1.228 & 0.5 & $60.57 \pm 3.42$ \\
1.249 & 0.5 & $56.34 \pm 3.22$ \\
2.026 & 1.5 & $19.05 \pm 0.99$ \\
2.078 & 1.0 & $15.35 \pm 0.81$ \\
3.179 & 3.0 & $6.19 \pm 0.34$ \\
4.017 & 2.5 & $4.28 \pm 0.22$ \\
5.027 & 3.0 & $3.24 \pm 0.18$ \\
6.185 & 3.0 & $3.53 \pm 0.19$ \\
7.122 & 2.0 & $3.85 \pm 0.25$ \\
9.148 & 2.0 & $3.16 \pm 0.18$ \\
11.011 & 2.0 & $3.1 \pm 0.18$ \\
13.087 & 2.0 & $2.94 \pm 0.16$ \\
29.078 & 2.0 & $3.14 \pm 0.26$ \\
43.924 & 2.0 & $2.23 \pm 0.15$ \\
73.058 & 2.0 & $2.7 \pm 0.51$ \\
\hline
\end{tabular}
\end{table}

\subsection{Swift}
A full description of the event as seen with the different instruments on Swift at hard X-ray,
soft X-ray, and UV/optical wavelengths will appear in Kowalski et al. (2014 in prep.)
Swift light curves were generated for the event using the on-line tool \url{http://www.swift.ac.uk/user_objects/}. Inspection of the
light curve reveals a main event, followed by a secondary event approximately 10$^{4}$s after the initial trigger. 
Numerous lower-level enhancements are present following these in the light curve for $\sim$20 days after the trigger. 
This is biased by the irregular observing cadence, which after a day introduces
large data gaps.  Figs 1 \& 2 show the light curve in the 0.3-10 keV energy band
obtained using Windowed-Timing (WT) and Photon-Counting (PC) observing modes. 
WT mode is used for high count rates and has a high-time resolution capability (2.2ms) 
with one dimensional position information and spectroscopy. 
PC mode is used for low count rates and retains
2-dimension imaging capability, spectroscopic capability, with minimum 2.5s time resolution.
The observing mode changes depending on the count rate of the source.

We estimated the energy conversion factor (ECF) for the Swift XRT during the interval 0.86-1.33 days after the trigger
by extracting a spectrum corresponding to this time interval and fitting it with a two temperature APEC model.
The integrated flux in the 0.3-10 keV band is 9.6$\times$10$^{-12}$ erg cm$^{-2}$ s$^{-1}$, corresponding to an average
of 0.31 counts s$^{-1}$ in the interval during which the spectrum was accumulated, or an ECF of 1.2$\times$10$^{30}$ erg count$^{-1}$. 
This suggests that the small X-ray flare seen at T0+$\sim$0.9 day has a peak $L_{X}$ of (2-3)$\times$10$^{30}$ erg s$^{-1}$.
Assuming that the ECF at early times in the flare is approximately the same, the peak $L_{X}$ is 10$^{32}$ erg s$^{-1}$.

\section{Discussion}
X-ray superflares, in which the soft X-ray luminosity increases by more than two orders of magnitude compared to `quiescent' levels, 
are relatively rare. 
Radio superflares have been seen in long timescale radio surveys, 
like the Green Bank Interferometer (GBI) which collected
$\sim$ daily monitoring of the radio fluxes of active binary systems, and they 
constrain the presence
of radio flares lasting for days, but typically RS~CVn and Algol systems
(Mutel et al. 1998; Richards et al. 2003) are targetted -- these are 
extreme examples of magnetic activity due to the effects of  tidal locking enforcing fast rotation in a
binary consisting of at least one evolved component.
M dwarf radio flare monitoring is usually of shorter duration (few hours at most),
and the AMI dataset discussed here occupies a unique niche in terms of radio variability monitoring.  The
extreme youth of the system adds another dimension to stellar parameter space which has not been systematically
explored at radio wavelengths. 

\begin{figure*}
\centerline{\epsfig{file=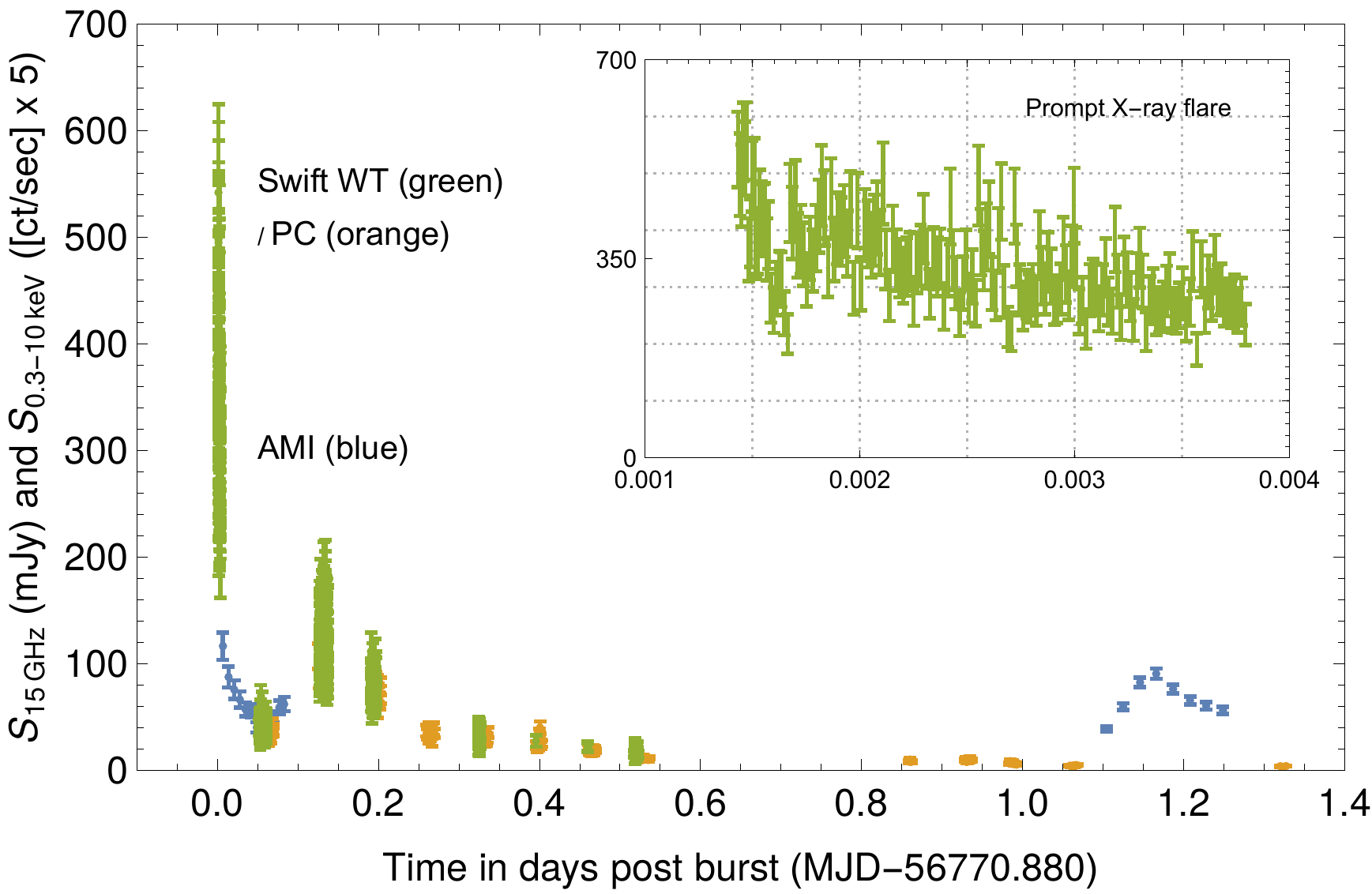, angle=0, width=14.5cm}}
\caption{As Fig 1, but plotted on a linear scale over the first 1.4 days post-burst. 
The inset shows the first set of X-ray data, between 0.002 -- 0.004 days.
Both the initial X-ray and radio flares appear to have peaked before the {\em Swift} and AMI observations began, at 124 and 318 seconds respectively. 
The radio flare between about 1.1--1.3 days post-burst fell in a gap between the {\em Swift} observations. }
\end{figure*}

The radio luminosity of this event eclipses even the 
large radio flare on the nearby flare star EV~Lac studied by Osten et al. (2005), by about an order of magnitude, and is probably the most luminous incoherent radio flare ever observed from a M-dwarf (brighter flares seen at lower frequencies are coherent in origin).
The peak radio flux from DG~CVn of $\approx$ 100 mJy implies a luminosity of 4.5$\times$10$^{16}$ erg s$^{-1}$ Hz$^{-1}$, which is about 20 times higher
than the large EV~Lac radio flare. While such extreme flares are commonly seen on active binary systems, the binary nature
of this system likely does not influence the magnetic activity: the separation of 0.2" at 18 pc implies a separation of
3.6 AU, or $\approx$2500 R$_{\star}$, and so the two stars are not magnetically interacting. Instead, the youth of the
system is the most likely explanation.  Bower et al. (2003) noted a large mm outburst from a K type star
in the Orion Nebula Cluster, with a peak radio flux density of $\approx$160 mJy, which assuming the star is single and at the
distance of the ONC ($\sim$450 pc), suggests an even higher L$_{r}$ of 4$\times$10$^{19}$ erg s$^{-1}$ Hz$^{-1}$. 

The detailed timing of the radio flaring compared to the X-ray flaring is shown in Figs 1 \& 2.  We explore two methods to assess the relation of the radio flare to the X-ray flaring.
The first uses the relationship established by G\"{u}del \& Benz (1993, hereafter GB93) between X-ray and radio luminosities of a large
sample of objects, including active M dwarfs, $L_{X}/L_{R}$=10$^{15.5\pm0.5}$Hz.  
The measurements were made at a slightly higher
frequency than those in GB93. However, radio spectra of active stars are generally flat in the microwave range, and so using the 15 GHz radio luminosity without frequency correction is sufficient for rough comparison.
The radio and X-ray data overlap at only a few times during the $\sim$20 days of the combined monitoring: 
at around 0.055 days post-trigger, the X-ray luminosity is $\sim$8.7$\times$10$^{30}$ erg s$^{-1}$, and 
the radio luminosity is 1.5$\times$10$^{16}$ erg s$^{-1}$ Hz$^{-1}$,  with a ratio $L_{X}/L_{R}$ of 6$\times10^{14}$ Hz.
The peak radio luminosity 
of the main flare is $\sim$ 4.5$\times$10$^{16}$ erg s$^{-1}$ Hz$^{-1}$, and that of the secondary event occuring
$\sim$ 1 day after the Swift trigger is $\sim$3$\times$10$^{16}$ erg s$^{-1}$ Hz$^{-1}$.
The $L_{X}/L_{R}$ ratio is $\sim$10$^{14}$ for the secondary peak, demonstrating that while this flare had a peak radio luminosity comparable to the prompt event, it was considerably less X-ray loud. The stellar data in GB93 were not, however, taken during flares (and generally not simultaneously) so the comparison of flare peaks is perhaps not appropriate.

Another relationship expected between signatures of particle acceleration and those of plasma heating during a flare is the Neupert effect (Neupert 1968). Under the standard flare scenario, the particle acceleration happens during
the initial impulsive phase of the flare:
the nonthermal particles deposit their energy in the lower stellar atmosphere, resulting in plasma heating and consequent X-ray radiation from the thermal plasma.
This predicts that the radio flare should {\it precede} an X-ray flare. We note that while stellar flares have been observed which do obey the
Neupert effect (Hawley et al. 1995, G\"{u}del et al. 1996, 2002, Osten et al. 2004), the opposite sense  (in which a radio flare occurs after an X-ray flare) have also been observed (Osten et al. 2005, Bower et al. 2003). For the prompt event, it is not really possible to tell whether there is any lag between the X-ray and radio emission, or vice versa, since the rise phase of both
X-ray and radio events were missed.
For the secondary radio flare around 1.1 days, there is a hint in the X-ray light curve of some weak rebrightening a few hours before the flare, but the link is inconclusive. It is clear, though, that during this event the source has a much larger ratio of radio to X-ray emission than during the prompt flare.

The utility of the radio light curve is apparent in inspecting the behavior in Figure~1. There is a large gap in Swift light
curve points between the initial $\sim$ 300 s of data and the next set of data points, more than 1 hour later.
The radio data confirm that during this time there were no secondary events.  
The nonthermal electrons acclerated during the flare event are thought to deposit their energy lower in
the stellar atmosphere, where they are collisionally stopped by the higher
column density. There will be temporal evolution of the energy
distribution of electrons, due to collisional and radiative losses, which is difficult to analyze with only a single 
frequency. 
Previous multi-frequency radio flares have shown a changing spectral index during the flares:
$\alpha$ (where $S_{\nu} \propto \nu^{\alpha}$) increases during the rise phase, then decreases during the decay
from its maximum value at the flare peak (Osten et al. 2005, Osten et al. 2004).  

Approximately 5 days after the trigger, the radio light curve shows a distinct flattening at a flux density of 2--3 mJy, which may indicate the return to the quiescent level of emission. This is consistent with flux levels observed on previous 
occasions at longer wavelengths:  it is a source in the FIRST catalog (Becker, White \& Helfand 1995) at 20 cm with a flux density of
2.2 mJy, and was detected on two occasions in Bower et al. (2009, where it is listed as GJ 3789) at 6 cm with flux densities of 6.2 and 3.1 mJy. This also suggests that the timescale for the radio flare is $\sim$4 days. Such a long timescale outburst
is unusual for active stars (although not for close binaries such as RS CVn variables), and this long duration may instead signal an association with one of the earlier, longer duration X-ray flares. 

\subsection{Summary}

In a very rapid robotic response to a {\em Swift} gamma-ray trigger, we have measured a prompt radio flare associated with an X/gamma-ray outburst from the nearby active binary DG CVn, as well as a secondary radio flare about one day later. These data are unique in probing the very early phases of radio emission from a stellar superflare, and should prove invaluable in testing and developing our models for particle acceleration in such systems. From an experimental point of view, they demonstrate the technical feasibility and high potential scientific impact of robotic response modes for radio telescopes, something which we strongly believe should be incorporated into the design of the Square Kilometre Array. 

\section*{Acknowledgements}

We thank the staff of the Mullard Radio Astronomy Observatory for
their invaluable assistance in the operation of AMI. Andy Beardmore
provided valuable assistance with the {\em Swift} data. The anonymous
referee provided pertinent and prompt comments which improved the 
paper. This project was largely funded by European Research Council Advanced Grant
267697 ``4 Pi Sky: Extreme Astrophysics with Revolutionary Radio 
Telescopes''.

\end{document}